\documentclass[fleqn,twoside]{article}
\usepackage{espcrc2}
\usepackage{epsf}

\usepackage{graphicx}
\usepackage[figuresright]{rotating}

\newcommand{\AmS}{{\protect\the\textfont2
  A\kern-.1667em\lower.5ex\hbox{M}\kern-.125emS}}
\newcommand{\be}{\begin{equation}}
\newcommand{\ee}{\end{equation}}
\newcommand{\ba}{\begin{eqnarray}}
\newcommand{\ea}{\end{eqnarray}}
\def\vx{\vec{\mbox{$x$}}}
\def\vk{\vec{\mbox{$k$}}}
\def\v0{\vec{\mbox{$0$}}}

\title{$SU(2)$ Running Coupling Constant and Confinement \\
       in Minimal Coulomb and Landau 
       Gauges\thanks{Talk presented by A.\ Cucchieri.}}

\author{Attilio Cucchieri\address[Bras]{IFSC-USP, Caixa Postal 369,
                                    13560-970 S\~ao Carlos, SP,
                                    Brazil}\thanks{Research partially
        supported by FAPESP, Brazil (Project No.\ 00/05047-5).},
        Tereza Mendes\addressmark[Bras]$^{\dag}$
        and
        Daniel Zwanziger\address{Department of Physics, NYU, New York,
                              USA}\thanks{Research partially supported
                              by the National Science Foundation,
                              grant no.\ PHY-0099393.
                              \vspace*{0.1cm}}}

\begin{document}

\begin{abstract}
We present a numerical study of the space-space and time-time components
of the gluon propagator at equal time in the {\it minimal Coulomb} gauge,
and of the gluon and ghost propagators in the {\it minimal Landau} gauge.
This work allows a non-perturbative evaluation of the {\it running
coupling constant} and a numerical check of Gribov's {\it confinement}
scenarios for these two gauges. 
Our simulations are done in pure $SU(2)$
lattice gauge theory at $\beta = 2.2$. We consider several lattice volumes
in order to control finite-volume effects and extrapolate our results to
infinite lattice volume.
\end{abstract}

\maketitle

\section{INTRODUCTION}

An essential step for understanding and extracting physical information
from gauge theories is the elimination of redundant gauge degrees of
freedom. This is usually done by choosing a representative on each
orbit of gauge-related fields ({\it gauge fixing}). In Ref.\
\cite{gribov} Gribov showed that the Coulomb and Landau gauge-fixing
conditions do not fix the gauge fields uniquely, i.e.\ there are
many gauge-equivalent configurations satisfying the Coulomb or Landau
transversality condition. These {\it Gribov copies} do not affect
perturbative calculations, but their elimination could play a crucial
role for non-perturbative features of gauge theories, such as color
confinement and hadronization.

\setcounter{footnote}{0}
In order to get rid of the problem of spurious gauge copies, Gribov
restricted \cite{gribov} the physical configuration space to the
region $\Omega$ of transverse configurations,
for which the Faddeev-Popov operator
is non-negative.\footnote{In the Coulomb case the transversality
condition and the positiveness of the Faddeev-Popov operator are
satisfied on each time slice.
}
This region is delimited by the {\it first Gribov horizon}, defined
as the set of configurations for which the smallest, non-trivial 
eigenvalue of the Faddeev-Popov operator is zero.
We now know that $\Omega$ is {\it not} free of Gribov copies and that
the physical configuration space has to be identified with the {\it
fundamental modular region} \cite{fmr,fmr2}. Nevertheless, the
region $\Omega$ is of interest in numerical simulations, since 
it is the space of configurations satisfying the usual lattice
Coulomb or Landau gauge condition.

The restriction of the path integral, which defines the partition
function, to the region $\Omega$ implies a {\it rigorous} inequality
\cite{vanish} for the Fourier components of the gluon field. From this
inequality, which is a consequence only of the positiveness of the
Faddeev-Popov operator, it follows that the region 
$\Omega$ is bounded by a certain ellipsoid $E$.
This bound causes a strong suppression of the (unrenormalized)
transverse
gluon propagator $D^{\rm tr}$ in the infrared limit
\cite{gribov,vanish}. More precisely, it was proven \cite{vanish} that,
in the infinite-volume limit, $D^{\rm tr}$ {\it
vanishes} at zero momentum, although the rate of approach to $0$, as a
function of the momentum or of $L$, was not established.
This is in marked contrast to the {\it divergence} in
the infrared limit of the free massless propagator.

Finally, because of entropy considerations \cite{gribov,critical}, the
Euclidean probability gets concentrated near the first
Gribov horizon\footnote{This has been verified numerically in the
minimal Landau gauge \cite{fmr3}.} where the inverse
of the Faddeev-Popov matrix
diverges. This causes an enhancement of the ghost propagator $G(k)$ in
the infrared limit \cite{gribov,fmr2}. This enhancement is a clear
indication of a long-range effect in the theory that may result in
color confinement.

The confinement scenario is particularly simple
in the minimal Coulomb gauge where the ghost propagator determines
directly the Coulomb interaction \cite{gribov,coul}. In fact, in
this case, confinement of color, i.e.\ the enhancement at
long range of the color-Coulomb potential $V(R)$, is due to the
enhancement of $G(\vk)$ at small momenta.
At the same time, the disappearance of gluons from the
physical spectrum is manifested by the {\it suppression} at
$\vk = \v0$ of the propagator $D_{ij}(\vk, k_4)$ of
3-dimensionally transverse would-be physical gluons.
Remarkably, $V(R)$ is the
instantaneous part of the $4$-$4$ component of the gluon propagator
$D_{44}(\vx, t)$, and is a renormalization-group-invariant
quantity \cite{coul}.
Its Fourier transform $\widetilde{V}(\vk)$ may serve to define the
running coupling constant of QCD by considering
$x_0 g_{C}^2(|\vk|) = \vk^2 \widetilde{V}(\vk)
= g^{2}_{0}\,\vk^2\,D_{44}(\vk)$,where
$x_0 = 12 N/(11N - 2N_f)$ has been calculated in \cite{rgcoul}.
Clearly, if
the color-Coulomb potential $V(R)$ is governed by a string tension
at large distances, i.e.\ $\widetilde{V}(\vk)$
goes like $1 / \vk^{4}$ at small momenta, then
$g_{C}^2(\vk) \sim 1 / \vk^2$ in the infrared limit.

Similarly, in Landau gauge, one can consider \cite{alphaL} the running
coupling constant $ g^{2}_{L}(k) = g^{2}_{0}\,[\,k^2\,D(k)\,]
\,[\,k^2\,G(k)\,]^2 $.
This is also a renormalization-group-invariant quantity since (in
Landau gauge) $Z_g \, Z_3^{1/2}\, \widetilde{Z_3} \,=\,\widetilde{Z_1} 
\,=\,1 $.
In this case we obtain $g^2_L (k) \sim k^{-2}$ if, for
example, $D^{tr}(k) \sim \mbox{const}$
and $G(k) \sim k^{-4}$ in the infrared limit. On the contrary,
if the gluon progator goes to $0$ in the infrared limit and
the ghost propagator blows up not faster than $k^{-4}$
then $g^{2}_{L}(k)$ has an
infrared fixed point \cite{alphaL}.

\vspace*{0.1cm}
We test these theoretical predictions with data from
a numerical study of $SU(2)$ lattice gauge theory, without quarks, in
the minimal Coulomb and Landau gauge at $\beta = 2.2\, $. Simulations 
were done at different lattice volumes $L^4$, in order to check for 
finite-size effects and, if possible, to extrapolate to infinite
lattice volume. 
Details of notation and numerical simulations are given,
for the Coulomb case, in \cite{numcoul} and will be presented, for the
Landau case, in \cite{future}.
For these simulations we have used a cluster of ALPHA work-stations
(Coulomb and Landau data) and a PC cluster (Landau data) at the 
Dept.\ of Mathematical Physics of the University of S\~ao Paulo
(DFMA/USP).\footnote{We thank Jorge L.\ deLyra for kindly providing
us with access to these clusters.} In the Landau case we have also 
used a PC cluster\footnote{16 nodes and a server with 866MHz 
Pentium III and 256/512 MB RAM memory, operating
with Debian Linux.} at the Institute of Physics of the University
of S\~ao Paulo, S\~ao Carlos (IFSC/USP).

\begin{figure}[hbt]
\begin{center}
\vspace*{-3.4cm}
\epsfxsize=0.45\textwidth
\leavevmode\epsffile{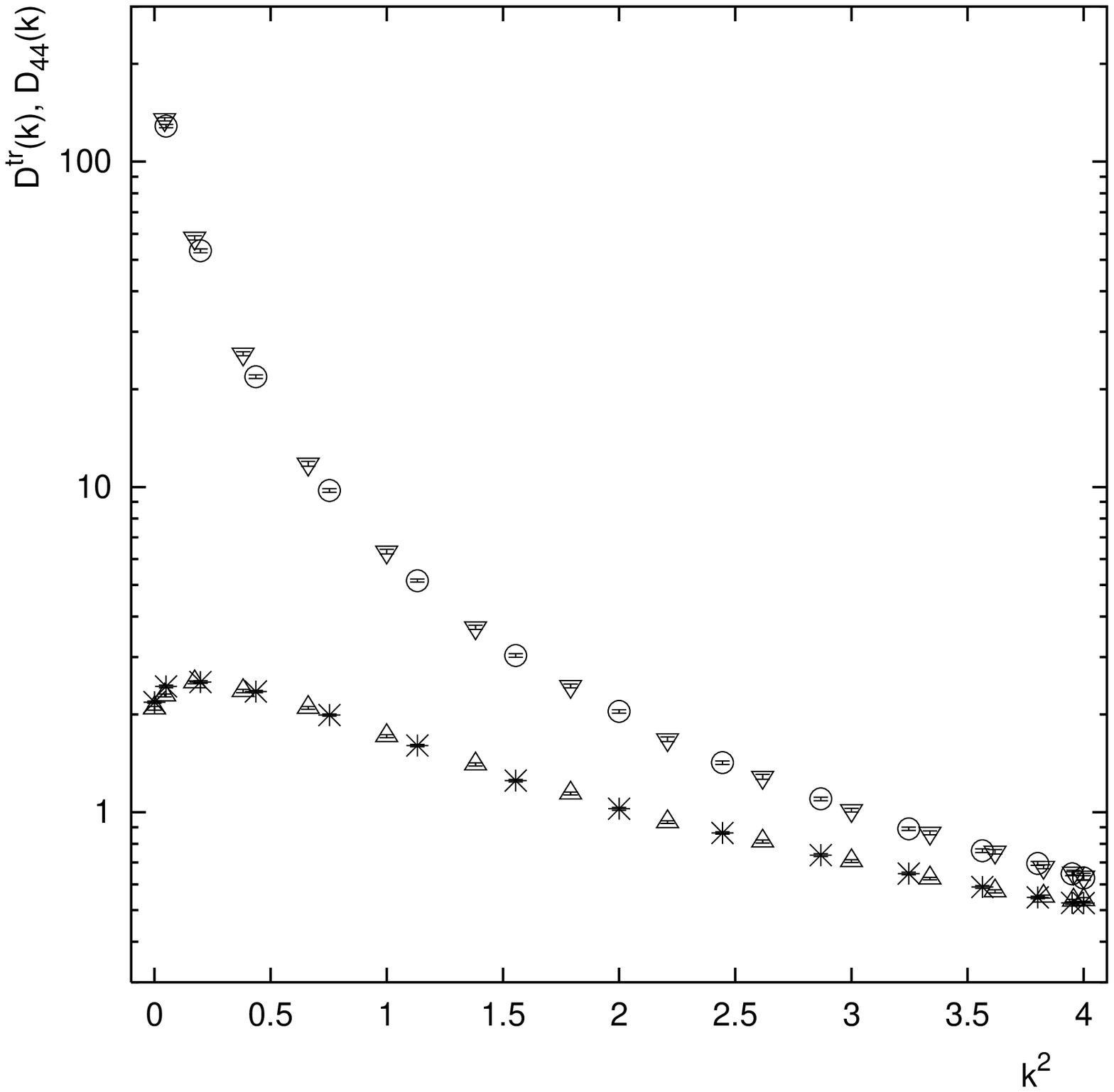}
\vspace*{-0.8cm}
\caption{Plot of the gluon propagators
$D^{\rm tr}(\vk)$ (lower curve) and
$D_{44}(\vk)$ (upper curve) as a function
of the square of the lattice momentum $\vk^2$ for
$L = 28$ (symbols $\ast$ and $\bigcirc$ respectively) and
$L = 30$ (symbols $\triangle$ and $\bigtriangledown$ respectively).
Notice the logarithmic scale on the $y$ axis.
\vspace*{-0.8cm}}
\label{fig:coulomb}
\end{center}
\end{figure}

\begin{figure}[hbt]
\begin{center}
\vspace*{-3.4cm}
\epsfxsize=0.45\textwidth
\leavevmode\epsffile{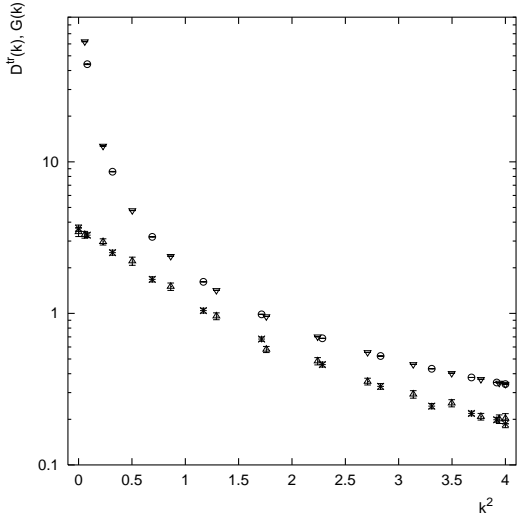}
\vspace*{-0.8cm}
\caption{Plot of the gluon propagator
$D^{\rm tr}(k)$ (lower curve) and of the ghost propagator
$G(k)$ (upper curve) as a function
of the square of the lattice momentum $k^2$ for
$L = 22$ (symbols $\ast$ and $\bigcirc$ respectively) and
$L = 26$
(symbols $\triangle$ and $\bigtriangledown$ respectively).
Notice the logarithmic scale on the $y$ axis.
\vspace*{-0.8cm}}
\label{fig:landau}
\end{center}
\end{figure}

\section{RESULTS}

Our data in the minimal Coulomb gauge \cite{numcoul,numcoul2} clearly show
(see Fig.\ \ref{fig:coulomb}) that the equal-time transverse
gluon propagator $D^{\rm tr}(\vk, L)$ passes through
a maximum and decreases as the momentum $\vk$ approaches $\v0$ (for fixed
$L$), and that $\widetilde{V}(\vk)$ is more singular than
$1/{\vk}^2$ at low $\vk$, which indeed corresponds to a long-range
color-Coulomb potential.
We have also obtained \cite{numcoul2}
an excellent 2-pole fit for $D^{\rm tr}(\vk,
L)$. Our fit indicates that the poles occur at complex $ m^2 = x(L) \,
\pm \, i y(L)$. In the infinite-volume limit we have $m^2 = 0 \, \pm \,
i y$, for $y = 0.375 \pm 0.162$ in lattice units, or $y = 0.330 \pm 
0.142$~GeV$^2$ for the location of the gluon poles in ${\vk}^2$.
(It follows from the Nielsen identities \cite{nielsen}
that these poles are independent of the gauge parameters.)

Similarly, in Landau gauge (see Fig.\ \ref{fig:landau}), the
transverse gluon propagator $D^{\rm tr}(k, L)$ is suppressed in the
infrared limit, while the ghost
propagator $G(k)$ is more singular than $1/k^2$.
Also here a 2-pole fit can probably be used to
fit the data for $D^{\rm tr}(k, L)$. In this case,
however, we need to improve the statistics in order to have better
control over the fits, and we need to simulate at larger lattice volumes
(we went up to $26^4$) to probe the infinite-volume limit.

Finally, for the running coupling constant $g^2$ we consider \cite{numcoul}
the fitting formula $ k^2 = \Lambda^2 \ \exp[(b g^2)^{-1}]
 \ [(b g^2)^{-r} + z  (b g^2)^{\alpha}]^{-1} \ , $
which implicitly defines $g^2$. Here $r = 102/121$, and
$ \Lambda, b, z$ and $\alpha$ are fitting parameters.
For small $g^2$, which corresponds to large momenta,
this formula is dominated by the first term in the denominator,
whereas for large $g^2$, i.e.\ small momenta,
it is dominated by the second term in the denominator.
In particular, $\alpha$ governs the strength of the singularity of
$g^2$ in the infrared limit; the case $g^2 \sim 1 / k^2$
corresponds to $\alpha = 1$.

In the Coulomb case we obtain \cite{numcoul}
$\alpha = 1.9 \pm 0.3$; this corresponds to
$g_{C}^2 \sim 1/|\vk|$, and
$V(\vk) \sim 1/|\vk|^3$ at low momentum.
In the Landau case we have
$\alpha = 2.5 \pm 0.3$, which implies
$g_{L}^2 \sim 1/ k^{0.8}$.
In both cases we have fitted the data with
a low momentum cut-off at ${\vk}^2 = 0.5$.
Of course, an
extrapolation in $\beta$ will be necessary to
determine the strength of these singularities in the continuum limit.

\end{document}